# Drainage of foam films

Roumen Tsekov
Department of Physical Chemistry, University of Sofia, 1164 Sofia, Bulgaria

A brief review of the classical theory of foam film drainage is presented as well as a new theory accounting for the effect of thickness non-homogeneities in the film developed by the film drainage.

The science of foams is as old as classical mechanics since the first systematic study in this area has been carried out by Newton [1]. He has noticed that soap films, losing liquid in time, pass through various thicknesses with characteristic colors. When the thinning process advances black spots appear which are sometimes unstable and the film ruptures at their place. These structures are nowadays known as Newton's black films. During the next centuries the interest to thin liquid films (TLFs) has increased [2-4] due to their importance for disperse systems such as foams, emulsions and suspensions widely employed in technology. The stability of dispersions is directly related to the process of thinning and rupture of TLFs dividing bubbles, droplets and particles [5]. At present, intensive investigations are performed on foam films formed from aqueous solutions of surfactants with biological origin. Special attention [6] is paid to TLFs stabilized by phospholipides because of the important role of these substances in the structure of biological membranes and metabolism of the living cells. The simplest TLF system, a single microscopic foam film, has proven to be a useful tool for model investigations. In many aspects the results of such a study are applicable to other types of TLFs as well, e.g. emulsion films, films on solid and liquid substrata. The contemporary versions of the interferomertic method proposed by Scheludko and Exerowa [7] enable measurement of the local kinematics in time of the film thickness. It is a challenge presently for theoreticians to supply a detailed theory of the TLF drainage. Scheludko [8] has also proposed a formula for the thinning rate of horizontal microscopic films formed in the center of a double concave drop

$$V = \frac{2h^3 \Delta p}{3\eta R^2} \equiv V_{\text{Re}} \qquad (1)$$

where $V$ is drainage velocity, $h$ and $R$ are the thickness and radius of the film, respectively, $\eta$ is the dynamic viscosity and $\Delta p$ is the difference between the capillary and disjoining pressures. Equation (1) is analogous to an expression known as the Reynolds law [9] for the velocity of approach of two parallel rigid discs separated by a liquid slit. The necessary conditions for applicability of the Reynolds law to foam films are tangentially immobile plane-parallel film surfaces and equilibrated Plateau border. In practice, however, many deviations from Eq. (1) have been observed and the reason should be found in violations of these compulsory conditions.

Deviations from the Reynolds law due to tangential surface mobility have been experimentally established [10-12] and theoretically described [13-15]. The usual way to produce relatively stable axisymmetrical films is to stabilize them with surfactants. Since the film drainage causes an interfacial gradient of the surfactant adsorption, a gradient of the surface tension appears and the corresponding Marangoni force slows down the interfacial flux [14-16]. The theory predicts a thinning rate satisfying Eq. (1) again but with a reduced viscosity $\bar{\eta} = \eta/(1+3\eta D/aE)$, where $a$, $E$ and $D$ are the surfactant adsorption length, Gibbs elasticity and diffusion coefficient, respectively. This means that $V$ is again inversely proportional to the square of the film radius $R$. Many experimentally determined rates of thinning of foam and emulsion films, however, exhibit much weaker dependence on the film size: $V$ is inversely proportional to the film radius $R$ to the power of about 0.8. Such a behavior cannot be explained by surface mobility, the effect of which is not expected to alter the functional dependence of the thinning rate on the film radius. Moreover, tangential immobility on the film surfaces can be achieved at higher surfactant concentrations in the film: for $aE \gg 3\eta D$ the film surfaces become immobile and $\bar{\eta} = \eta$. Malhotra and Wasan [17] have proposed a model explaining the deviations of the $R$-dependence of $V$ from the Reynolds law by disturbances of the equilibrium state in the meniscus due to liquid outflow from the film. The latter are confirmed by exact numerical calculations [18], too. This effect diminishes, however, when the volume of the meniscus is much larger than the volume of the film and the film drainage is relatively slow. In addition, the relaxation in the Plateau border is as quicker as deeper it is. Therefore, for sufficiently thin films with sharp Plateau borders one can expect that the meniscus remains quasi-equilibrated during the film drainage.

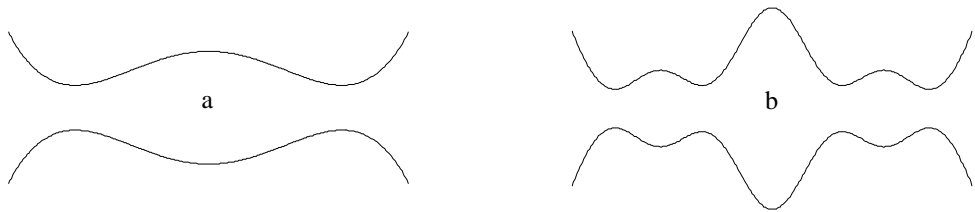

**Fig. 1** Film shape during the film drainage: (a) single and (b) multiple axisymmetrical dimples

Obviously, the thickness non-uniformity of the films is the basic reason for deviations from Eq. (1) in the case of TLFs stabilized by surfactants. The influence of the surface deforma-bility on film thinning is the most complicated effect. Equation (1) requires drainage of the liquid to be strictly symmetrical with respect to the center of the circular film confined between two parallel flat surfaces. With solid surfaces this condition is provided by their rigidity while the profile of a film with fluid surfaces is determined by the uniform external pressure upon the film area and capillary and surface forces. Many experimental investigations [19-22] have shown that for large radii the film surfaces are not strictly planar and parallel, i.e. the films are non-homogeneous in thickness. Due to the film geometry, the liquid in the film center flows initially slower as compared to that in the film periphery. In this way a characteristic shape is formed which is known under the name of "dimple" (see Fig. 1a). Ruckenstein and Sharma [23] have put forward an explanation of the deviations from Eq. (1) through the peristaltic action of surface hydrodynamic propagating waves formed in the process of the film thinning. Observations show, however, quasi-static surface corrugations rather than running surface waves.

The theoretical problem of dimple formation and its evolution has been widely studied [24-29]. The outflow from TLFs is described by the Navier-Stokes equations in the frames of the Reynolds lubrication approximation valid if $h \ll R$ [3]. For an axisymmetrical dimple, the corresponding continuity and dynamic equations in cylindrical coordinates are $\partial_\rho(\rho u) + \rho \partial_z w = 0$ and $\partial_\rho P = \eta \partial_{zz} u$ where $u$ and $w$ are the radial and normal components of the hydrodynamic velocity, $P$ ($\partial_z P = 0$) is the pressure in the film, and $\partial$ indicates partial derivatives. Integrating the dynamic equation twice on $z$ with the boundary condition of tangentially immobile film surfaces $u(z = \pm H/2) = 0$, one has $8\eta u = (4z^2 - H^2)\partial_\rho P$ where $H$ is the local film thickness. Substituting now $u$ in the continuity equation and integrating once more under the symmetry condition $w(z=0) = 0$, the normal component of the hydrodynamic velocity can be expressed by the relation $24\eta\rho w = \partial_\rho[\rho z(3H^2 - 4z^2)\partial_\rho P]$. Finally, employing this equation and the kinematics relation $\partial_t H = 2w(z = H/2)$ yields a differential equation governing the evolution of the local film thickness in time

$$12\eta\rho\partial_t H = \partial_\rho(\rho H^3 \partial_\rho P) \tag{2}$$

The further analysis requires a relationship between the film pressure and the local film thickness. It is supplied by the normal force balance on the film surfaces

$$P = G - \sigma\partial_\rho(\rho\partial_\rho H)/2\rho - \Pi(H) \tag{3}$$

where $G$ is the constant pressure in the gas phase, $\sigma$ is the film surface tension and $\Pi$ is the disjoining pressure. The capillary term in Eq. (3) is one for relatively flat film surfaces.

In general, the evolution of the film thickness is governed by Eqs. (2) and (3) and a set of relevant boundary conditions. There are three main problems, however, to calculate rigorously the film profile evolution. First, Eq. (3) requires an expression for the dependence of the disjoining pressure from the film thickness. Apart from the classical van der Waals and electrostatic components of $\Pi$ [3] many other interesting effects of macroscopic interaction have been discovered. Here one can mention hydrophobic, structural, protrusion, undulation, etc. forces. In the case of micellar and vesicular suspensions the disjoining pressure exhibits a periodic dependence on $H$ which is responsible for the film stratification. Second, Eqs. (2) and (3) are non-linear and one may expect a chaotic behavior of their numerical solutions. Indeed, it is the conclusion of many experiments that the film evolution is very sensitive to the manner of film preparation, i.e. the initial conditions are important. Unfortunately, the latter are not strictly specified from an experimental point of view. Third, the rigorous boundary conditions could be defined only far away from the film in the meniscus where Eqs. (2) and (3) are no longer valid owing to violation of the lubrication approximation. For this reason some approximate conditions suggested by the experiments are employed, e.g. minimum of the film shape at the border and constant thinning rate, which are questionable and less general.

Let us consider now the derivation of Eq. (1) and, hence, the range of validity of the Reynolds law to foam films. Taking the film surfaces to be flat the local thickness $H$ in the film is everywhere equal to the average thickness $h$. In this case Eq. (2) reduces to $12\eta\rho V = -h^3 \partial_\rho (\rho \partial_\rho P)$ where $V = -\partial_t h$. Integrating twice this equation under the constrain that the pressure at the firm rim is equal to the pressure $M$ of the liquid in the meniscus, one obtains $P = M + (R^2 - \rho^2)3\eta V / h^3$. Now, combining this result with Eq. (3) and integrating once again, one gets

$$\partial_\rho H = \rho \Delta p / \sigma - \rho(2R^2 - \rho^2)3\eta V / 2\sigma h^3 \qquad (4)$$

where $\Delta p \equiv G - M - \Pi(h)$. As seen in Fig. 1a, the film shape exhibits minimum at the film border. Hence, $(\partial_\rho H)_R = 0$ and it follows immediately from Eq. (4) that the rate of drainage of the film obeys the Reynolds law (1). Integrating Eq. (4) once again and remembering that $h$ is the average film thickness, one gets the film profile

$$H = h + (2R^4 - 6R^2\rho^2 + 3\rho^4)\Delta p / 12\sigma R^2 \qquad (5)$$

As is seen, Eq. (5) describes a dimple with a thicker part in the film center and a thinner ring at the film border as is presented in Fig. 1a. Of course, the film is not more planar and to obtain the range of validity of the Reynolds law one should require at least a positive thickness in the film thinnest part, i.e. the barrier rim. Using Eq. (5) one can transform the corresponding inequality $H(R) \geq 0$ to

$$R \leq \sqrt{12\sigma h / \Delta p} \qquad (6)$$

Therefore, the Reynolds law is valid only for small films with radii restricted by inequality (6). This is exactly the conclusion of many experiments which clearly show that the Reynolds law is only applicable to sufficiently small films which are relatively homogeneous in thickness. For instance, Radoev et al. [20] have reported that films with $h$ = 250 nm, $\Delta p$ = 35 N/m$^2$, $\sigma$ = 35 mN/m and $\eta$ = 1 mNs/m$^2$ obey Eq. (1) if they are smaller than 0.05 mm which coincides with the value of the transition radius predicted by Eq. (6). Note that all the films approach their equilibrium state with the Reynolds rate of thinning since in the limit $\Delta p \to 0$ inequality (6) is fulfilled for arbitrary film radii. This is not surprising because the equilibrium foam films are flat.

The question is what happens with TLFs with radii out of the range of inequality (6). The experimental results [20] indicate an increase of the thickness non-homogeneity in amplitude and number with increasing film size. This corresponds to increasing divergence from the idealized models of planar film profile and single axisymmetric dimple to formation of several centers of concurrent thinning in the film. The single dimples in such films are not stable and break spontaneously down to more complex structures, an ensemble of several smaller films (see Fig. 1b). The minimum entropy production theorem says that this should result in a more favorable hydrodynamic regime and, therefore, in an increased rate of drainage. The existence of multiple dimples schematically presented in Fig. 1b was recently demonstrated experimentally [30] in large TLFs. This new phenomenon was theoretically explained as a result of loss of dynamic correlation in the behavior of the local parts in the film [31]. The effect of thickness non-homogeneity in the multiple dimple on the TLF drainage was also studied both experimentally [30] and theoretically [31]. A new formula describing the thinning rate of dimpled thin liquid films was derived

$$V = \frac{1}{6\eta} \sqrt[5]{\frac{h^{12} \Delta p^8}{4\sigma^3 R^4}} \qquad (7)$$

In contrast to the Reynolds law, the predicted rate of drainage in Eq. (7) is inversely proportional to the 4/5 power of the film radius which is in excellent agreement with the experi-mentally determined functional dependence of $V$ vs. $R$ (see Fig. 2). The theory predicts the number of the film thickness non-homogeneity $n = (R^2 \Delta p / 16\sigma h)^{2/5}$, too. As is seen, the number of corrugations increases simultaneously with the film radius and driving pressure as required by the experiments. It is easy to show now that the thinning rate of large TLFs $V = n\sqrt{n}\, V_{Re}$ is always larger than $V_{Re}$ and reduces to the Reynolds one only for films with radii given by Eq. (6) since $n = 1$.

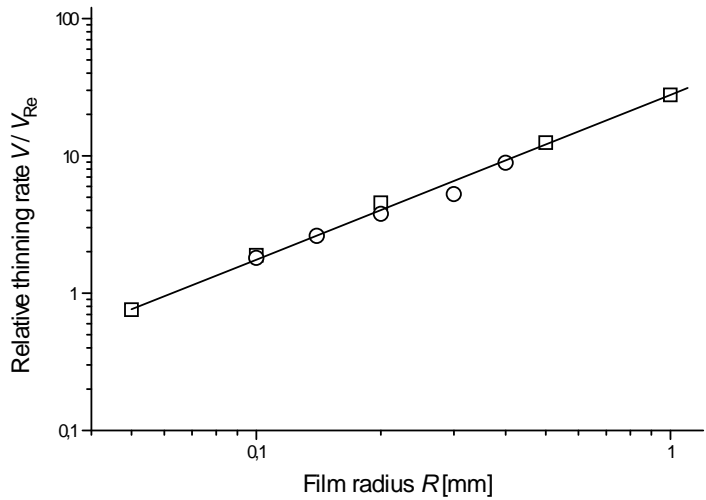

**Fig. 2** Dependence of film thinning rate on the film radius: squares [20] and circles [32] represent experimental results, the line is the prediction of Eq. (7) for films with $h$ = 250 nm, $\Delta p$ = 35 N/m$^2$, $\sigma$ = 35 mN/m and $\eta$ = 1 mNs/m$^2$. The slope of the line is 6/5

The theoretical models describing the film profile as axisymmetrical dimple take into account the deformation of the film surfaces during the thinning process. Sometimes, however, these models do not correspond well to the real film shapes, especially in large TLFs formed by quick initial withdrawal of liquid. More complex phenomena take place in such films [33] and the cylindrical symmetry of the thinning process, assumed in the present consideration, is violated [34]. As a rule, the asymmetric film drainage is much quicker than the axisymmetric one due to the peristaltic pumping effect [23, 33]. Owing to the stochastic nature of the film shape in this case, the description of the film drainage is more complicated and may require a statistical approach.